\documentclass[11pt]{article}

\usepackage[utf8]{inputenc}
\usepackage[T1]{fontenc}
\usepackage[english]{babel}

\usepackage{amsmath, amssymb}

\usepackage{booktabs}

\usepackage{graphicx}
\usepackage{caption}

\usepackage{xcolor}
\usepackage[colorlinks=true,linkcolor=blue,citecolor=blue,urlcolor=blue]{hyperref}

\usepackage{geometry}
\usepackage[round,authoryear]{natbib}
\usepackage{lineno}
\usepackage{doi}

\newcommand{\rtp}{(r,\theta,\varphi)}
\newcommand{\tp}{(\theta,\varphi)}
\newcommand{\smp}{\sin(m \varphi)}
\newcommand{\cmp}{\cos(m \varphi)}

\newcommand{\Plm}{P_{\ell}^m(\cos\theta)}

\newcommand{\glm}{g_{\ell}^m}
\newcommand{\hlm}{h_{\ell}^m}

\newcommand{\qlm}{q_{\ell}^m}
\newcommand{\qlmj}[1]{q_{\ell,#1}^m}
\newcommand{\slm}{s_{\ell}^m}
\newcommand{\slmj}[1]{s_{\ell,#1}^m}

\newcommand{\blm}{B_{\ell m}}
\newcommand{\ylm}{Y_{\ell}^m}
\newcommand{\vlm}{V_{\ell m}}

\newcommand{\velmm}{V_{\ell m}^{E-}}
\newcommand{\velmp}{V_{\ell m}^{E+}}

\newcommand{\vilmm}{V_{\ell m}^{I-}}
\newcommand{\vilmp}{V_{\ell m}^{I+}}

\newcommand{\glmmag}{g_{\ell,{\mbox{\sc \scriptsize Mag}}}^m}
\newcommand{\hlmmag}{h_{\ell,{\mbox{\sc \scriptsize Mag}}}^m}

\newcommand{\lmax}{\ell_{\text{max}}}

\newcommand{\bnabla}{\boldsymbol{\nabla}}
\newcommand{\rv}[1]{\textcolor{black}{#1}}
\newcommand{\rvv}[1]{\textcolor{black}{#1}}

\newcommand{\supportinginfoTitle}{
    \begin{center}
        { \textit{\textbf{\textcolor{lightgray}{Accepted for publication in Geophysical Research Letters}}}}\\[1.5em]
        {\Large \textbf{Motional induction in Ganymede's ocean}}\\[1em]
        \textbf{Simon Cabanes}$^{1}$, \textbf{Thomas Gastine}$^{1}$, and \textbf{Alexandre Fournier}$^{1}$ \\[0.5em]
        \small $^{1}$Universit\'e Paris Cit\'e, Institut de physique du globe de Paris, UMR 7154 CNRS, 1 rue Jussieu, F-75005 Paris, France\\[1em]
        Corresponding author: Simon Cabanes (cabanes@ipgp.fr)
    \end{center}
    \vspace{2em}
}

\newcommand{\AF}[1]{\textcolor{black}{#1}}

\begin{document}

\selectlanguage{english}

\supportinginfoTitle

\vspace{-0.5em}
\today

\section*{Abstract}
We investigate the magnetic signature of oceanic circulation in Ganymede's subsurface ocean 
using kinematic induction modeling. 
 Our approach couples zonal jet flows
from rotating thermal convection simulations with magnetic
field models incorporating Ganymede's internal dynamo and external contributions
from Jupiter. 
 We solve the induction equation in spherical geometry for
deep-ocean (493~km) and shallow-ocean (287~km) scenarios
with varying magnetic Reynolds numbers. 
 Ocean flows generate
a predominantly toroidal magnetic field through the omega-effect,
with a weaker poloidal component pervading beyond the conductive ocean layer.   
 For some, but not all, induction configurations,  analysis  \rv{of the time-averaged Lowes-Mauersberger} spectra  reveals that ocean-induced signals dominate
at spherical harmonic degrees $\ell \geq 4$. Deep ocean scenarios with magnetic Reynolds numbers
above unity produce surface magnetic signals up to 9~nT. 
Our results demonstrate that Ganymede's intrinsic magnetic field creates favorable conditions
for detecting subsurface ocean dynamics,  
thus emphasizing the need for low-altitude
orbits for the Juice probe.

\section*{Plain Language Summary}

\rv{Jupiter’s moon Ganymede harbors a vast subsurface ocean. Understanding its dynamics is key to assessing how its circulation may facilitate the exchange of heat and materials essential for life between the moon’s interior and the overlying ice shell.}
However, since this ocean is buried under 
kilometers of ice, direct observation of this circulation appears 
to be impossible.

Our study shows that Ganymede's ocean currents might actually 
reveal themselves through magnetic signals. Ganymede 
is unique among icy moons because it has its own magnetic field, 
generated deep in its metallic core. When electrically conducting
ocean water flows through this magnetic field, it creates additional 
magnetic signatures that could be detected by spacecraft magnetometers.

Using computer simulations, we find that strong east-west ocean
currents may generate magnetic signals up to 9 nanoteslas 
at Ganymede's surface. The European Space Agency's Juice mission, 
en route to study Jupiter's moons, should be capable of 
detecting these signals when it orbits Ganymede, 
with enhanced sensitivity if low-altitude orbits are performed. 
 
 This research suggests that magnetic measurements could provide 
a welcome glimpse into the hidden dynamics of an extraterrestrial 
ocean, opening a new window into worlds that might support life.

\section{Introduction}
Data from the Galileo and Juno missions, along with observations from the Hubble Space Telescope, provide compelling evidence that Ganymede possesses important prerequisites to be considered 
habitable. 
The presence of salts, sulfur compounds, and organic matter on its frozen surface, coupled with 
signs of tectonic activity and cryovolcanism, collectively suggests the existence of a well-mixed, 
liquid-water ocean beneath the ice crust, \rvv{potentially overlying high-pressure ice layers \citep{lucchita80,mccord01,patterson10,tosi23,journaux20,kalousova2025}}. 
Electromagnetic methods also played a crucial role in detecting this subsurface ocean. In 
particular, magnetic observations of Jupiter’s time-varying magnetosphere, as it diffuses through 
the moon’s interior, revealed induction signals and damped auroral oscillations, both indicative of 
a global, electrically conductive ocean 
\citep{Khurana98,zimmer00,Schilling07,Saur10,Seufert11,saur15,vance21}.
\rvv{Taken together, these observations broaden the range of potentially habitable environments to include icy worlds.}

Among the pivotal questions arising, a crucial inquiry concerns how oceanic flows support the exchange of heat and materials essential for life between the planet’s interior and the upper ocean/ice interface. 
These exchanges are enhanced by convective motions \cite[e.g.,][]{Soderlund14, Vance05} or by mechanically driven flows induced by tides, libration, and orbital precession \cite[e.g.,][]{Lemasquerier17, Rovira19, Tyler08}—mechanisms thought to be energetic enough to mix ocean waters and shape thermal and compositional gradients that may support life.  
So far, insights into these ocean dynamics have come exclusively from numerical simulations, which consistently show the emergence of global east–west zonal (axisymmetric) jets under the influence of planetary rotation \citep{soderlund19,bire22,kvorka22,kvorka24}.
However, such simulations significantly overestimate viscous dissipation, making them highly sensitive to the choice of boundary conditions (e.g., free-slip vs. no-slip), and preventing a clear consensus on the expected structure of these zonal jets. 
This limitation highlights the need for \rv{observational} constraints on ocean dynamics, 
\rv{a need that upcoming missions such as ESA’s Juice \citep{vanhoolst24} and NASA’s Europa Clipper \citep{pappalardo24} will help to address.}

Despite the long-lived scientific interest of these subsurface oceans, a signal directly 
emanating from the ocean dynamics remains to be identified.
Only information derived from ice-penetrating radar ($\sim$ 10 km penetration depth), \rv{geodetic} 
measurements of the ice crust, and, potentially, observations of cryovolcanism at the surface are anticipated to deliver insights into the underlying dynamics \rv{\citep{grasset13,vanhoolst24}}. 
\rv{\citet{tyler11}, \citet{vance21} and \citet{Sachl25} }
proposed that Europa (another Galilean moon), which also harbors a subsurface ocean with active oceanic flows, could generate a magnetic signal through its interaction with Jupiter’s ambient magnetic field.
However, this signal is expected to reach only $\sim$1 nT, near the lower detection limit of Europa Clipper’s magnetometers.
Unlike Europa, Ganymede has an intrinsic dynamo producing a strong magnetic dipole
\citep{kivelson96}, 
which will enhance \rv{motional} induction \rv{ in its ocean \cite[e.g.][]{vance21,kvorka26}}. 
This parallels Earth’s magnetic environment, where tidally-driven magnetic signals have been detected in satellite data \citep{tyler03,grayver19}.
However, Ganymede’s deep, global-scale oceanic circulation may amplify magnetic induction even further compared to Earth’s more limited oceanic flows.

The goal of this study is to assess \rv{the imprint} of this induction mechanism \rv{on time-averaged Lowes-Mauersberger (LM henceforth) spectra \citep{lowes74,mauersberger56}}.

\section{Ocean-Induction setup}\label{sec:Ocean-Induction setup}
\subsection{Theoretical framework}\label{sec:Theoretical framework}
We solve the induction equation in Ganymede's ocean,
\begin{equation}\label{eq:induction}
\frac{\partial \mathbf{B}}{\partial t} = \nabla \times (\mathbf{U} \times \mathbf{B}) + \frac{1}{\mu
\sigma} \nabla^2 \mathbf{B},
\end{equation}
where $\mathbf{B}$ is the magnetic field, $\mathbf{U}$ is the prescribed
ocean flow, $\mu$ is the magnetic
permeability and $\sigma$ is the electrical conductivity, \rv{assumed to be homogeneous}. 
This kinematic approach neglects 
the feedback of the magnetic field onto the ocean flow. If $\mathbf{B}(\mathbf{U})$ denotes
the solution to Eq.~\eqref{eq:induction}, we define the field 
$\mathbf{B}_{\mathrm{mi}}$ 
due to motional induction
in Ganymede's ocean by 
\begin{equation}
\mathbf{B}_{\mathrm{mi}}\equiv   \mathbf{B}(\mathbf{U}) - \mathbf{B}(\mathbf{U=\mathbf{0}}),  
\label{eq:bmi}
\end{equation}
which amounts to subtracting the solution obtained for a motionless, yet conductive,  
ocean. 
The ability of $\mathbf{U}$ to generate a detectable $\mathbf{B}_{\mathrm{mi}}$ 
depends
on the balance between the inductive and diffusive terms in Eq.~\eqref{eq:induction}. This 
balance is
quantified by the magnetic Reynolds number 
\begin{equation}
R_m \equiv U D \mu \sigma, 
\label{eq:rm}
\end{equation}
where $U$ represents a typical
flow velocity and $D$ is the ocean thickness. Therefore, predictions of  $\mathbf{B}_{\mathrm{mi}}$ 
require knowledge of $D$, $\sigma$, and $U$, all of which are impacted by uncertainty. 

Models of Ganymede’s interior indicate that the moon is a differentiated body, consisting of a metallic core of 
radius $r_c \gtrsim 500$ km, surrounded by a silicate mantle and a $\sim 800$~km-thick (ice–liquid) water layer above~\citep{Anderson96,Kuskov01,Sohl02}. 
\rv{Values for ocean thickness $D$ and electrical conductivity $\sigma$ rest 
on the thermodynamically consistent water‐layer models of \citet{vance18}, which}
incorporate realistic electrical conductivity profiles along adiabatic temperature gradients. 
Their results suggest that multiple pressure-induced phase transitions 
occur within the water layer, leading to an ocean 
embedded between two distinct ice layers.
These models also account for the thermodynamic properties of pure water and aqueous magnesium sulfate (MgSO$_4$), the latter being considered the dominant salt in Ganymede’s ocean~\citep{kargel91}.

From the ensemble of models proposed by \citet{vance18}, we selected 
two end-member scenarios assuming a 10 wt\% MgSO$_4$ solution \cite[following the synthesis by][]{soderlund19}. 
These are summarized in Table~\ref{tab:scenarios}. 
\begin{table}[h!]
\centering
\caption{Summary of key parameters for the deep and shallow ocean scenarios.
The radius of Ganymede $r_g$ is set to 2631~km.
The  magnetic permeability $\mu$ is that of vaccum, $\mu = 4 \pi \times 10^{-7}$ H/m. For a given geometry, 
the low estimates of flow velocity $U$ and magnetic Reynolds number $R_m$ correspond to dissipation controlled by turbulent 
quadratic drag at the ocean-ice interface, while the high estimate assumes that this dissipation is due to linear Ekman friction.} 
\label{tab:scenarios}
\begin{tabular}{lrrr}
\hline
\textbf{Parameter} & unit & \textbf{Deep-ocean} & \textbf{Shallow-ocean} \\
\hline
Depth $D$ &  km & 493 & 287 \\
Ice thickness $D_I$ & km &  26 & 95 \\lll
Radius ratio $r_i/r_o$ & - & 0.811 & 0.887 \\
Water density $\rho$ & kg/m$^{3}$ & $1.24 \times 10^{3}$ & $1.23 \times 10^{3}$ \\
Electrical conductivity $\sigma$ & S/m & 4 & 2.5 \\
Buoyancy power $P$ & m$^2$/s$^3$ & 30.11 & 16.22 \\
Flow velocity $U$ (quadratic drag) & m/s & 0.053  & 0.036  \\
$R_m^{\textsc{q}} = U D \mu \sigma$ (quadratic drag) & -  & 0.13  & 0.032  \\
Flow velocity $U$ (Ekman friction) & m/s &  0.71 &  0.43 \\
$R_m^{\textsc{e}} = U D \mu \sigma$ (Ekman friction) & -  & 1.77 &  0.39 \\
\hline
\end{tabular}
\end{table}
Each scenario is assigned a typical geostrophic \rv{(i.e. invariant along the direction of planetary rotation)} flow velocity $U$
by balancing the available power per unit mass $P$ with the boundary-layer dissipation rate $\mathcal{F}_J(U)$ at the ice–ocean interface \citep{jansen23}.
To account for this dissipation, \citet{jansen23} assume a parameterized turbulent quadratic drag\rvv{, leading to an energy dissipation} of the form
\begin{equation}
\mathcal{F}_J(U) \approx \frac{2 C_D U^3}{D},
\mbox{ (quadratic drag) }
\label{eq:quadra}
\end{equation}
with the turbulent drag coefficient $C_D = 5 \times 10^{-3}$ \citep{brenner21}. Alternatively, \citet{cabanes24} argue that dissipation is primarily controlled by Ekman friction at the outer boundary, which leads to  
\begin{equation}
\mathcal{F}_J(U) \approx \dfrac{(r_o \nu \Omega)^{1/2}}{\left(r_o^2 - s_{mid}^2\right)^{3/4}} U^2, \mbox{ (Ekman friction)}
\label{eq:linear}
\end{equation}
where $\Omega$ is the planetary rotation rate, $\nu$ is the kinematic viscosity of water, and the mid-radius $s_{mid} = (r_i + r_o)/2$, where $r_i$ and $r_o$ are the inner and outer radii of the ocean, respectively.
Assuming that ocean circulation is driven by thermal convection, two estimates of $U$ are obtained by equating  Eq.~\eqref{eq:quadra} or Eq.~\eqref{eq:linear}  
with the mean buoyancy power $P$
of the two models selected from \citet{vance18}, which leads to two 
values
of $R_m$, $R_m^{\textsc{q}}$ and $R_m^{\textsc{e}}$, respectively. Values of $U$ and $R_m$ are reported in Table~\ref{tab:scenarios} for both deep- and shallow-ocean scenarios.  When linear Ekman friction is assumed, $R_m \sim \mathcal{O}(1)$ indicates sizable  
magnetic induction from ocean flow. 

\subsection{Oceanic flow structure}\label{sec:Oceanic flow structure}
As in \citet{cabanes24}, we assume that the flow $\mathbf{U}$ is driven
by rapidly rotating thermal convection due to an adverse temperature contrast $\Delta T$ 
maintained  between the inner and outer radii of the shell, which rotates at angular velocity $\Omega$. 
The boundaries are rigid and isothermal. 
The fluid mechanical problem is governed by three dimensionless numbers: 
the Ekman number $E = \nu/\Omega D^2$, the Rayleigh number $Ra = \alpha g_o D^3 \Delta T/ \nu \kappa$, and the Prandtl number $Pr = \nu/\kappa$, where $\kappa$ is the thermal diffusivity, $\alpha$ is the thermal expanslivity, 
and $g_o$ is the gravitational acceleration at the outer boundary. We set $Pr=1$.
\citet{soderlund19} suggests Ekman numbers in the range $E \sim 10^{-10} - 10^{-13}$ and Rayleigh numbers $Ra \sim 10^{20} - 10^{24}$, indicating that the oceanic circulation falls within a dynamical regime moderately influenced by planetary rotation. Accordingly, we adopt the simulation by \citet{cabanes24} with $E = 10^{-5}$, $Ra = 2 \times 10^9$, and a radius ratio $r_i/r_o = 0.8$ to represent the deep-ocean scenario. To represent the shallow-ocean scenario, we 
conduct a new simulation using $E = 3 \times 10^{-5}$, $Ra = 1 \times 10^8$, and $r_i/r_o = 0.9$, a parameter choice constrained by the increased numerical cost associated with the larger radius ratio. 
\begin{figure}[h!]
\centering
  \includegraphics[width=0.7\textwidth]%
    {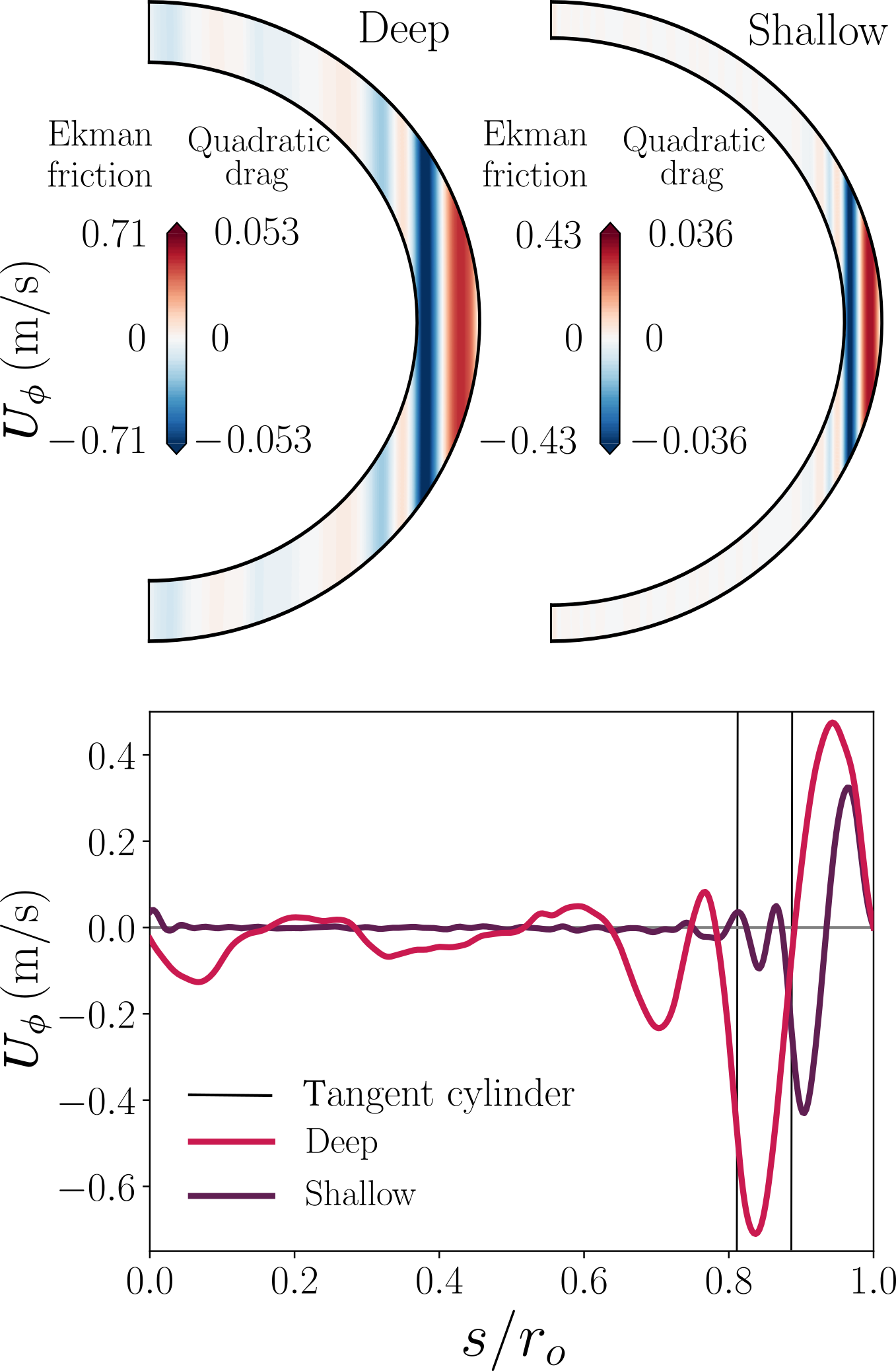}
    \caption{Top: Meridional cross-sections of the geostrophic zonal velocity ($U_\phi$) for the deep- and shallow-ocean scenarios. For each scenario, two scales are 
    provided, depending on the dissipation mechanism thought to 
    operate (see text for details). Bottom: geostrophic velocity
    versus normalized radius $s/r_o$\rvv{, assuming that
    dissipation is governed by Ekman friction}. The vertical line at $s/r_o=0.811$ (resp. $0.887$) marks
    the tangent cylinder for the deep- (resp. shallow-) ocean. 
    \label{fig:Mapsvelo}}
\end{figure}


Despite the gap between numerically accessible and planetary regimes, we assume that the meridional cross-sections of the flow, presented in Fig.~\ref{fig:Mapsvelo}, can be extrapolated to Ganymede’s ocean.  
\rvv{These maps have been slightly rescaled from \citet{cabanes24} to account for the small differences in the target radius ratios considered here (0.811 instead of 0.8, and 0.887 instead of 0.9 see Table~\ref{tab:scenarios}).}
In both cases, the oceanic flow is dominated by prograde (red) and retrograde (blue) steady zonal jets that form predominantly outside the tangent cylinder. These maps represent the geostrophic zonal component of the simulated flow  ($U_\phi$), believed to be the most energetic in Ganymede’s ocean 
\rv{\cite[non-zonal flows being an order of magnitude weaker, implying 
a proportionally lower $R_m$,][]{cabanes24}},
and for which velocity magnitudes are given in Table~\ref{tab:scenarios}.
The lower panel of Fig.~\ref{fig:Mapsvelo} shows the associated velocity profiles along the normalized cylindrical radius $s/r_o$. These extrapolated steady geostrophic flows $\mathbf{U}$ enter the induction equation \eqref{eq:induction}, where they interact with 
Ganymede’s magnetic field $\mathbf{B}$. 

\subsection{Model of Ganymede’s Magnetic Field}\label{sec:magneticmodels}
In addition to induction in Ganymede's ocean, we consider two sources to Ganymede's magnetic field: its core dynamo
and the Jovian magnetosphere. From the ocean perspective, the latter is an external source while the former is an internal source. 
In insulating regions, both sources can be described by a spherical harmonic (SH) expansion using Gauss coefficients
(see~Supporting~Information~S1.1 for details). 
Let $\ell_{\text{max}}^E$ and $\ell_{\text{max}}^I$ denote the truncation of the SH representation of the external and 
internal sources, respectively. 

We derive the external Gauss coefficients from Jupiter’s internal magnetic model of \citet{connerney22} and the Jovian magnetodisc model of \citet{Connerney20}.
Using NAIF-produced SPICE kernels\footnote{freely available at \url{https://naif.jpl.nasa.gov/naif/data.html}} \citep{acton96}, we construct magnetic field maps of the Jovian magnetosphere in a Ganymede-centered right-handed planetocentric  coordinate system for a representative time span of 46 days, corresponding to 156 Ganymede orbital periods.  
Time-dependent external Gauss coefficients are obtained through a least-squares fit of the sequence of maps at Ganymede's surface radius using $\ell_{\text{max}}^E = 1$.  
\rv{This external‐field model also served to benchmark our induction setup, as we successfully reproduced the analytical diffusive response of 
a static ocean ($\mathbf{U}=\mathbf{0}$) derived by \citet{zimmer00}.}
The detailed computation of this external model and its salient properties are described in Supporting Information~S1.2. 

The internal Gauss coefficients due to Ganymede's dynamo are based on models by \citet{Weber22}, \citet{plattner23}, and \citet{jia25}, which combine flyby  measurements by the Galileo and  Juno probes. 
As detailed in Supporting information~S1.3,  these models are truncated at SH  degree 2, a consequence of the paucity of data that makes it impossible to resolve smaller scales. 
In any event, those models suggest that the dynamo field is dipole-dominated, as supported 
by numerical dynamo simulations by 
\citet[][his Figure~8]{christensen15}\rv{, and that it is about ten times stronger 
than the external field at Ganymede's surface}. Christensen's simulations also suggest the energetic content of the non-dipole field at the top of the dynamo region can be considered flat to first order, up to SH degree~8. 
Consequently, our working assumption is that at the top of Ganymede's core ($r=r_c$), the non-dipole \rv{LM}  spectrum is flat 
(for practical purpose up to $\ell_{\text{max}}^I=10$), with an energetic content per degree matching that estimated for the quadrupole by \citet{Weber22}, \citet{plattner23}, or \citet{jia25}, downward-projected at $r=r_c=500$~km \citep{vance18}.  
 For each of the three field models considered, we generate 11 random realizations of internal
 Gauss coefficients up to $\ell_{\text{max}}^I$, subject to the condition that the energy contained in each SH degree $\ell \ge 3$ matches that contained in SH degree $2$, which gives rise to 33 sets of dynamo Gauss coefficients. 
 On the physical time span considered for this study (a few weeks), we further assume that the internal Gauss coefficients are constant, unlike their external counterparts  (see Supporting Information~S1.3).

\subsection{Numerical method}\label{sec:Numerical method}
The induction equation \eqref{eq:induction} is solved in a spherical shell using \texttt{MagIC}\footnote{freely available at \url{https://github.com/magic-sph/magic}} \citep{wicht02,gastine16}. 
The code rests on a pseudospectral approach, and it utilizes the \texttt{SHTns} library for spherical harmonic transforms \citep{schaeffer13}\footnote{freely available at \url{https://gricad-gitlab.univ-grenoble-alpes.fr/schaeffn/shtns}}.
Both $\mathbf{U}$ and $\mathbf{B}$ are decomposed into their poloidal and toroidal components, which are in turn 
expanded in spherical harmonics up to degree $\ell_{\max}=192$ in colatitude $\theta$, order $m_{\max}=10$ in longitude $\phi$, and in Chebyshev polynomials up to degree $N_r=97$ in radius $r$. Time integration is performed using the second-order Crank-Nicolson Adams-Bashforth scheme \cite[e.g.][]{glatzmaier84}. 
 
While $\mathbf{B}$ contains, by design, non-axisymmetric features up to SH order $m=10$,  
$\mathbf{U}$ is axisymmetric (recall section \ref{sec:Oceanic flow structure}), and 
setting $m_{\max}=10$ suffices. In contrast, setting $\ell_{\max}=192$ is necessary to
provide sufficient resolution in colatitude.  
The external and internal Gauss coefficients used to model Ganymede’s ambient magnetic field enter the magnetic 
boundary conditions prescribed at the top and bottom of the ocean, respectively; a derivation of these conditions is given in Supporting Information~S2. 

\begin{figure}[h!]
  \centering
  \includegraphics[width=\textwidth]{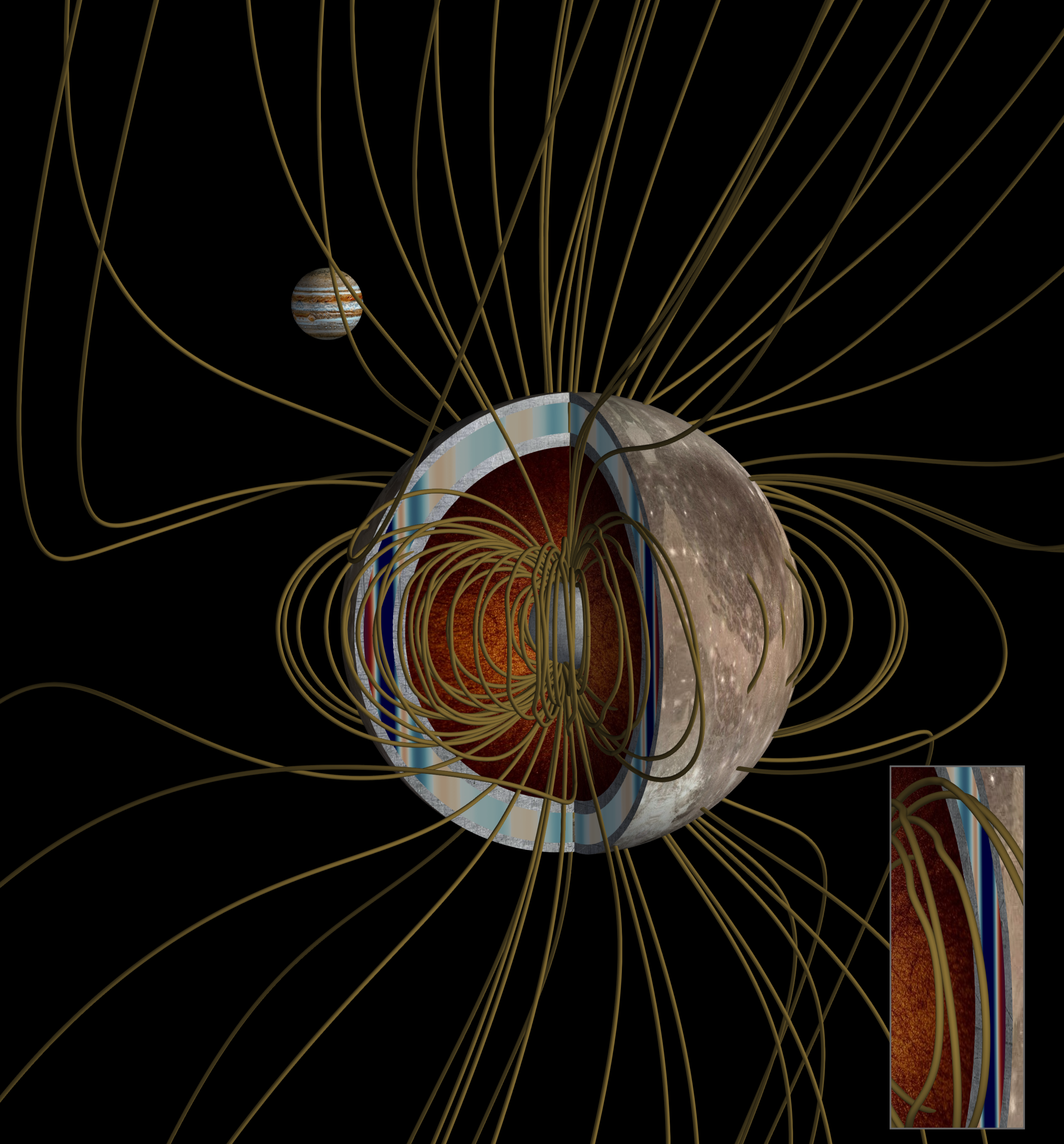}
    \caption{Numerical ocean-induction setup for Ganymede in a deep-ocean scenario with $R_m^{\textsc{e}}=1.77$. 
    Ganymede's internal structure is depicted, featuring the ice crust, the deep ocean with its simulated zonal 
    flow, the internal ice layer, the silicate mantle, and the metallic core. Yellow tubes visualize the magnetic field 
    lines.
    The bottom-right panel shows how magnetic field lines are sheared by the zonal (east–west) 
    jet flow in the ocean (blue/red denote westward/eastward jets, respectively), thereby reflecting the secondary 
    magnetic field induced by such dynamics. \label{fig:NumSetup}
    }
\end{figure}

We performed 132 numerical simulations, covering both deep- and shallow-ocean scenarios, two values of $R_m$ for each, and 33 realizations of the internal dynamo field. We ran the same number of simulations setting $\mathbf{U}=\mathbf{0}$, in order to compute $\mathbf{B}_\mathrm{mi}$ from Eq.~\eqref{eq:bmi}. Figure~\ref{fig:NumSetup} illustrates our approach with a deep-ocean example at $R_m = R_m^{\textsc{q}}=1.77$. The magnetic field lines highlight the three-dimensional structure of Ganymede’s ambient magnetic field. In the bottom-right inset, these field lines appear to be sheared by the ocean’s jet flow, revealing how such dynamics can generate a secondary magnetic field.

\section{Results}\label{sec:results}
\begin{figure}[h!]
  \includegraphics[width=\textwidth]%
    {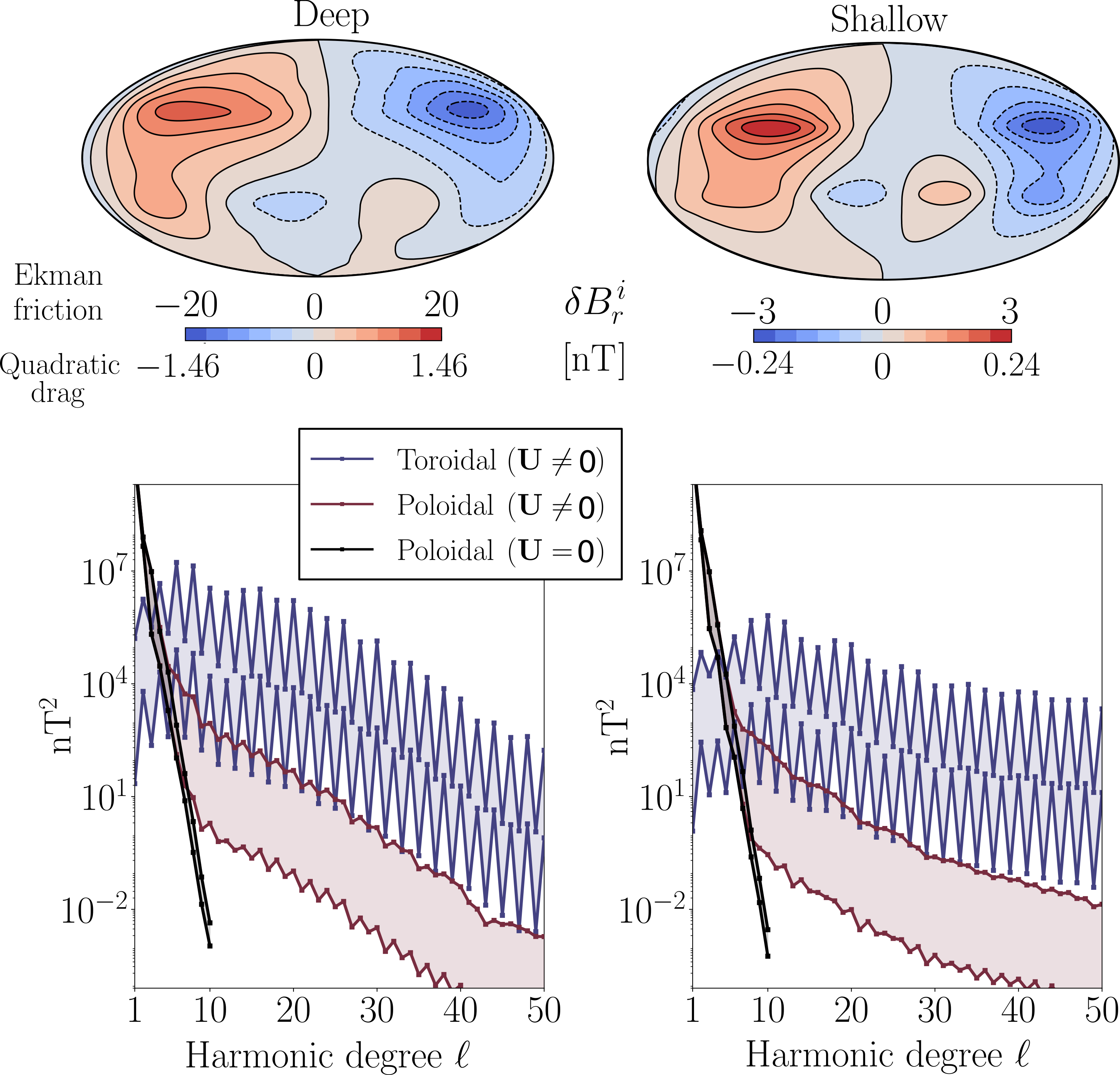}
  \caption{Top: Hammer maps of the radial
  component of the magnetic field induced by Ganymede's ocean zonal flow, $\delta B_r^i = \hat{\mathbf{r}} \cdot \mathbf{B}_{\mathrm{mi}}$, at the surface of the ocean $r=r_o$.  
  The bottom (resp. top) scale of the colorbars corresponds to $R_m=R_m^{\textsc{q}}$ (resp. $R_m=R_m^{\textsc{e}}$). 
Bottom: Spherical harmonic decomposition of toroidal and poloidal magnetic energy for the full range of tested magnetic Reynolds numbers. The spectra are time-averaged. Colored envelopes show energy with flow; black envelopes show energy without flow.
  } \label{fig:magspectra-poltor}
\end{figure}
Figure~\ref{fig:magspectra-poltor} \rv{(top)} shows maps of the ocean flow-induced 
radial magnetic field at the surface of the ocean $r=r_o$, 
defined as $\delta B_r^i = \hat{\mathbf{r}}\cdot \mathbf{B}_{\mathrm{mi}}$ \rv{(horizontal 
components are shown and discussed in Supporting Information~S5)}. 
The deep- and shallow-ocean scenarios generate non-axisymmetric magnetic fields, with ocean-induced signatures reaching 1.46~nT (resp. 20~nT) for the deep ocean, taking 
$R_m=R_m^{\textsc{q}}=0.13$ (resp. $R_m^{\textsc{e}}=1.77$),  and 0.24~nT (resp. 3~nT) for the shallow ocean with 
$R_m=R_m^{\textsc{q}}=0.032$ (resp. $R_m^{\textsc{e}}=0.39$).
\rv{These values are comparable in strength to those reported by \citet{kvorka26} and below the upper estimate of \citet{vance21}. }

Figure~\ref{fig:magspectra-poltor} \rv{(bottom)} presents the  time-averaged \rv{ poloidal and toroidal magnetic power} spectra (integrated over the oceanic shell) for the full set of simulations  (see Supporting Information~S4). 
In the motionless case ($\mathbf{U}=\mathbf{0}$, black envelopes), the magnetic field in the ocean is purely poloidal and dominated by the dipolar component ($\ell = 1$), with magnetic energy rapidly decreasing at higher harmonic degrees.
If $\mathbf{U} \neq \mathbf{0}$ instead (colored envelopes), oceanic motions substantially reinforce magnetic energy at higher degrees, and the resulting field becomes predominantly 
toroidal due to the azimuthal shearing of the ambient magnetic field by the zonal jets. 
This process is commonly referred to as the omega-effect \cite[e.g.][]{parker1955hydromagnetic}. 

However, this toroidal field is trapped inside the ocean layer. Consequently, when assessing whether a magnetic signal from ocean flow could be detected by spacecraft magnetometers orbiting above the ice shell, we focus on \rvv{its} poloidal component, which can pervade the region of space above the ocean. 
\begin{figure}[h!]
  \centerline{\includegraphics[width=\textwidth]{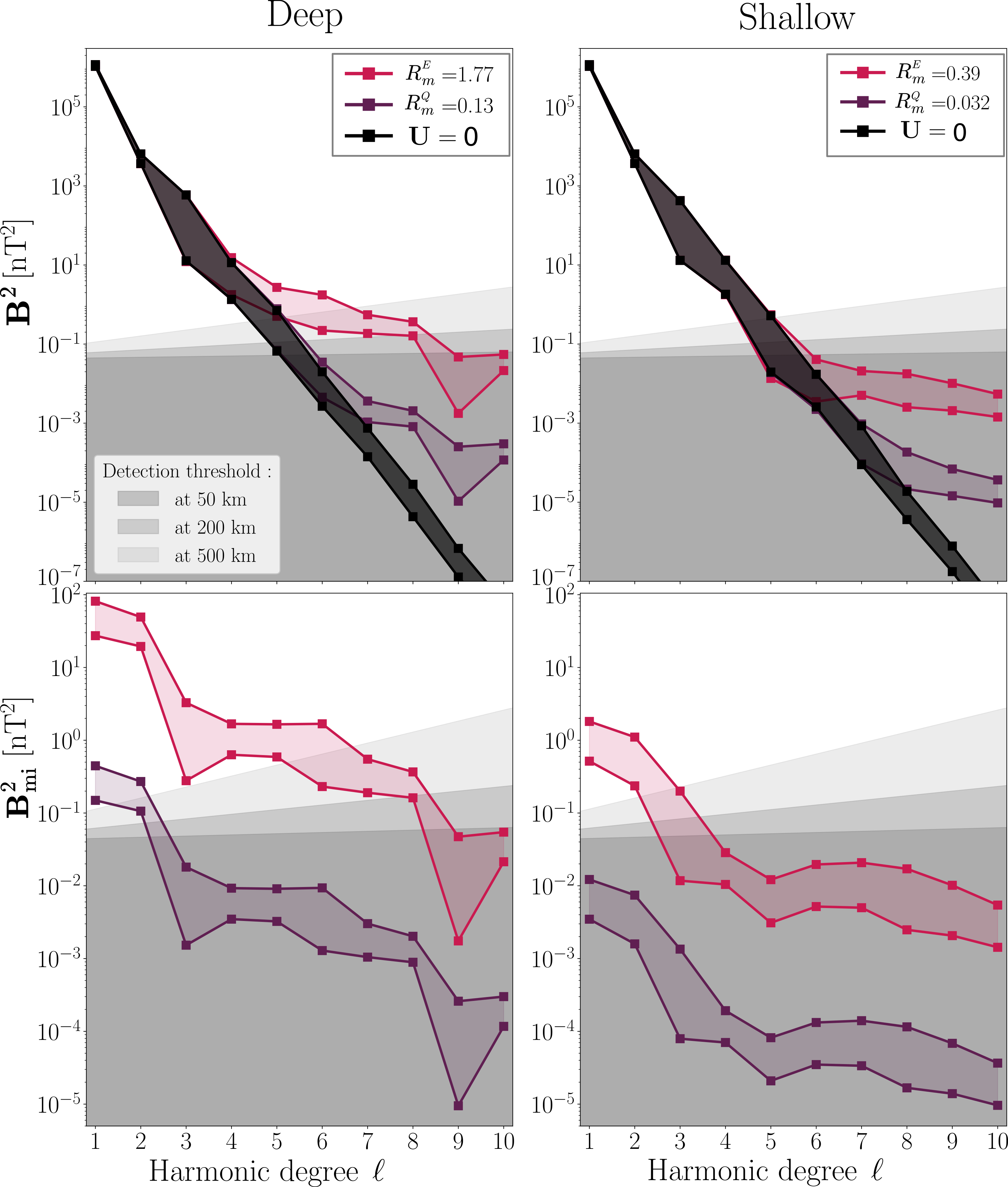}}
    \caption{
    Top: magnetic \rv{power} spectra at Ganymede's surface $r=r_g$, computed
    for the total magnetic field $\mathbf{B}$. 
    Bottom: magnetic \rv{power} spectra at $r=r_g$, due to motional induction in the ocean alone, computed for $\mathbf{B}_\mathrm{mi}$ (see text for details). All spectra are time-averaged. 
    \rv{
    The slope of the 0.2 nT detection threshold is caused by the amplification of measurement errors during downward continuation from spacecraft altitude to Ganymede's surface.}
    }\label{fig:MagSpectra}   
\end{figure}
Accordingly, Figure~\ref{fig:MagSpectra} presents time-averaged 
\rv{LM} spectra at Ganymede’s surface (for details consult Supporting Information~S3), separately for the deep- and shallow-ocean scenarios. 
The top panels of Figure~\ref{fig:MagSpectra} show the spectra \rvv{of $\mathbf{B}$} without oceanic flow as a black envelope, while those 
including ocean-induced magnetic fields for $R_m^{\textsc{q}}$ and
$R_m^{\textsc{e}}$ are displayed as colored envelopes.
In all cases, oceanic flow is responsible for a 
decrease of magnetic 
energy with SH degree $\ell$ \rv{less severe} than the one obtained with a motionless ocean. The latter corresponds to geometrical decay from the top of the dynamo region to the surface of Ganymede. 
 \rv{The break in trend due to the ocean-induced field is analogous to Earth's ``magnetic curtain'' \citep{roberts13}. On Earth, the large-scale field (low $\ell$) shows the geometrical decay of the dynamo field from the core, while a transition to a nearly flat spectrum occurs around $\ell \sim 13$, above which crustal magnetism dominates and conceals the small-scale geodynamo.}
 For Ganymede’s ocean-induced magnetic field, the transition occurs at degrees $\ell \geq 4$ or 6 in the deep-ocean scenario, and at $\ell \geq 6$ or 8 in the shallow-ocean scenario, depending on the value of $R_m$. This reflects that a deep ocean induces a larger-scale magnetic field
than a shallow ocean, whose zonal flows are confined  to a tighter latitudinal band
around the equator. 

Figure~\ref{fig:MagSpectra} also shows the detection threshold of 0.2~nT, characteristic of Juice’s magnetometers \rvv{\cite[on-ground fluxgate noise $<$0.2 nT at lower frequencies according to][see~Supporting~Information~S1.3 for details]{brown19}}. 
Therefore, magnetic signals with energies exceeding the grey areas will be detected by the mission during its orbit at altitudes of 500, 200, and optionally 50 km (not scheduled yet to our knowledge) 
above Ganymede’s surface.
For a deep ocean with $R_m=R_m^{\textsc{e}}=1.77$, the imprint of the ocean-induced magnetic signal significantly exceeds the threshold, whereas in the other configurations, the signal remains too weak for detection. 

Finally, the bottom panels of Fig.~\ref{fig:MagSpectra} show the 
energetic content of the poloidal component of $\mathbf{B}_\mathrm{mi}$
alone, again at $r=r_g$. In all induction scenarios but the shallow-ocean case with $R_m=R_m^{\textsc{q}}=0.032$, the ocean flow-induced magnetic field signal at large scales (i.e., $1 \leq \ell \leq 3$) exceeds the detection threshold. Over the full range of magnetic Reynolds numbers tested, the induced magnetic dipole reaches magnitudes between 0.06 and 9 nT, \rv{values that are comparable to the tidal-flow-induced magnetic signal ($\sim 1$ nT) measured above Earth's oceans by satellites \citep{tyler03,grayver19}}.
We stress that isolating the induced dipole component is contingent upon properly
separating the sources of Ganymede's magnetic field $\mathbf{B}$. 

In any event, a magnetic signal induced by oceanic flow is likely to be detected by Juice, except under the most pessimistic conditions involving a shallow ocean and a low magnetic $R_m$ approaching $R_m^{\textsc{q}}=0.032$, 
 assuming that turbulent quadratic drag is responsible for dissipation in 
 Ganymede's ocean.  
Detecting the large scales (dipole to octupole, say) of $\mathbf{B}_\mathrm{mi}$
is challenging in terms of the separation 
of sources, noting  that global coverage at low altitude can only help in this respect. 
 The exception is the more optimistic scenario, a deep ocean with $R_m \gtrsim1$, 
 in which $\mathbf{B}_\mathrm{mi}$ is readily detectable for $\ell \geq 4$, whereas the dynamo signal is, in contrast, considerably attenuated at those scales.

\section{Discussion}\label{sec:Discussio}
\rv{Our study investigates the potential for Juice's magnetic 
measurements to reveal Ganymede's subsurface ocean circulation. Using 
a kinematic induction framework, we assess the magnetic imprint of 
ocean flow. Given the uncertainties in key parameters -- ocean 
thickness, flow velocity, and electrical conductivity -- that control 
the magnetic Reynolds number ($R_m$), we performed 132 simulations 
spanning a broad range of plausible induction scenarios. 
These cover deep and shallow oceans, with thickness and conductivity 
from the thermodynamic models of \citet{vance18} and typical flow 
velocity estimates based on the energetic arguments of 
\citet{jansen23} and \citet{cabanes24}.}

Our simulations show that 
the interaction of oceanic zonal jets 
with Ganymede's dynamo field 
primarily generates a toroidal field via the omega-effect. 
Yet only the weaker extra 
poloidal component can permeate the insulating ice shell and beyond. 
This ocean-generated magnetic signal suffers significantly less geometrical attenuation than the dynamo-generated field originating $\sim$2000 km deeper in the core. At Ganymede's surface, this gives
rise to a distinct trend in the \rv{Lowes-Mauersberger} magnetic spectrum for spherical harmonic degrees  
$\ell \gtrsim 4$, 
which can be unambiguously attributed to ocean flow (recall  Figure~\ref{fig:MagSpectra}).

It is, however, noteworthy that only the most optimistic scenario with a deep ocean and $R_m=1.77$ induces a magnetic imprint strong enough to exceed the expected 0.2~nT detection threshold of Juice's magnetometers \rv{(bearing in mind that measurements may be 
impacted by additional uncertainty due to spacecraft attitude, range, and position)}. 
In all other scenarios, the ocean-induced signal remains detectable, but it is  confined to $\ell \leq 3$, where separating it from the dynamo contribution 
is challenging.  
This difficulty also highlights a key limitation of the internal magnetic field models  on which our study is based (see Section~\ref{sec:magneticmodels}). 
For this proof-of-concept exercise, we have indeed assumed that the internal dipole and quadrupole coefficients estimated by \citet{Weber22,plattner23,jia25} were entirely due to Ganymede's dynamo, whereas our analysis has shown that motional induction in the ocean could generate a few nanoteslas. This raises the crucial question of separability of sources. Addressing this issue will require future efforts to better constrain the specific dynamics of both Ganymede’s dynamo and oceanic flow, as well as their expected respective contributions to the magnetic spectrum observed at its surface.  Interestingly, \citet{otzen2024coestimation} recently showed in the context of geomagnetism how physics-informed prior information could be used to co-estimate two internal sources of Earth's magnetic field (in their case, the Earth's core and its crust). 

Nevertheless, a key result of this study is that a detectable magnetic signal exists in most induction scenarios, paving the way for probing Ganymede’s ocean circulation using Juice’s magnetic data. 
This opportunity relies heavily on the presence of Ganymede’s dipole-dominated
internal dynamo, which provides the background field upon which zonal jets act to generate a secondary magnetic signal. 
This rather unique situation contrasts with that of 
Europa, which lacks an internal dynamo. There, under the most favorable conditions 
identified by \citet{Sachl25}, namely, a 150~km thick ocean with an electrical conductivity of 
18~S/m and a 1~km thick ice layer, the ocean-induced magnetic field may reach 1~nT. This signal is approximately 
nine times weaker than the $\sim$9 nT induction maximum found in our most optimistic Ganymede scenario. 
\rv{This value is consistent with the $\sim$8 nT estimate of \citet{kvorka26}, making Ganymede the most promising target for detecting ocean flow via magnetic measurements, pending low-altitude Juice orbits. This conclusion could be mitigated by recent Juno-constrained analyses \citep{jia25}, which suggest a lower ocean conductivity ($\sigma \sim 0.1$ S/m), a thicker ocean ($D \gtrsim 100$~ km), and a $\sim150$~km thick ice shell.}

\rv{Our detectability assessment uses time-averaged Lowes-Mauersberger spectra to disentangle internal magnetic sources. Future work should analyze magnetic field maps at spacecraft altitude, following  
 \citet{kvorka26}. Plotting the magnetic energy per harmonic order $m$ could also quantify 
 the non-axisymmetric part of the dynamo field interacting
with the zonal jets, as illustrated in the similar, low-$R_m$ context of Jupiter by \citet{wicht24}.}
On the forward modelling side, future investigations should also include additional components of ocean circulation.
For example, \citet{cabanes24}  identified in some simulations planetary, non-axisymmetric, waves embedded within the equatorial zonal jets. The characterization of these 
waves and their potential for magnetic detection remain to be assessed.

\section*{Acknowledgments}
\rv{We thank the two anonymous reviewers for their constructive comments.}
Numerical computations were performed on the S-CAPAD/DANTE platform, IPGP, France.
The authors acknowledge the support of the French Agence Nationale de la Recherche
(ANR), under grant ANR-19-CE31-0019 (project revEarth).
This work has been funded by ESA, France in the framework of EO Science for Society, through contract 4000127193/19/NL/IA (SWARM + 4D Deep Earth: Core).

\section*{Conflict of Interest}
The authors declare no conflicts of interest relevant to this study.

\section*{Open Research}
Simulations were performed with the open-source \texttt{MagIC} code \citep{wicht02,gastine16}, available at \mbox{\url{https://github.com/magic-sph/magic}} (GPL license).
Simulation data and analysis scripts (version 1.0) are archived on Zenodo \citep{cabanes2025ganymede} \mbox{\url{https://zenodo.org/records/16612142}} and include magnetohydrodynamics outputs, magnetic field coefficients, Python processing scripts, and visualization workflows using \texttt{matplotlib} version 3.7.5 \citep{hunter07}. Figure 2 visualizations were generated with \texttt{ParaView} \citep{paraview}.

\section*{Supporting Information}  

\setcounter{section}{0}
\setcounter{figure}{0}
\setcounter{table}{0}
\setcounter{equation}{0}

\renewcommand{\thesection}{S\arabic{section}}
\renewcommand{\thefigure}{S\arabic{figure}}
\renewcommand{\thetable}{S\arabic{table}}
\renewcommand{\theequation}{S\arabic{equation}}

\section{Magnetic models}
\subsection{External and internal magnetic potentials}\label{Asec:ExtIntGC}
\begin{figure}[ht]
  \centerline{\includegraphics[width=0.6\textwidth]{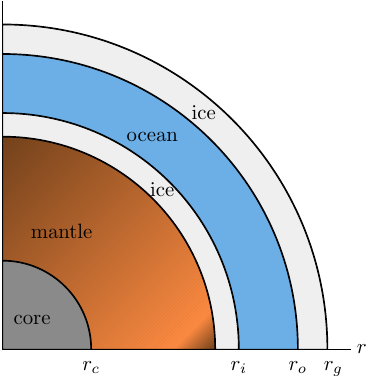}}
    \caption{Schematic spherically symmetric structure of Ganymede considered in this study. The mantle and ice layers are assumed to be electrically insulating. }
    \label{fig:coupeGany}  
\end{figure}

As illustrated in Fig.~\ref{fig:coupeGany}, we assume that 
the internal structure of Ganymede is spherically symmetric, characterized by nested layers of variable electrical conductivity: 
a conductive metallic core (radius $r<r_c$), a composite insulating layer consisting of a silicate
mantle and an ice layer ($r_c<r<r_i$), 
a conducting ocean ($r_i<r<r_o$), and an insulating ice cap ($r_o<r<r_g$). 

In a current-free region, the magnetic field $\mathbf{B}$ is a potential field, 
\begin{equation}\label{AEq:PF}
\mathbf{B} = - \bnabla V, 
\end{equation}
where $V$ is the magnetic potential. 
The potential is further decomposed  into an internal part $V^I$ and an external part $V^E$, corresponding to internal and external sources, respectively. 
In spherical geometry, $V^I$ and $V^E$  can be conveniently expressed using spherical harmonics, such that 
\begin{equation}\label{AEq:VIVE}
\begin{split}
V(r,\theta,\phi,t) &= V^I(r,\theta,\phi,t) + V^E(r,\theta,\phi,t)\\
 &= r_g \sum_{\ell=1}^{\ell_{\text{max}}^I} \sum_{m=0}^{\ell} \left( \frac{r_g}{r} \right)^{\ell+1} \left[\glm(t) \cmp + \hlm(t) \smp\right] \Plm\\
 &+ r_g \sum_{\ell=1}^{\ell_{\text{max}}^E} \sum_{m=0}^{\ell} \left( \frac{r}{r_g} \right)^{\ell} \left[\qlm(t) \cmp + \slm(t) \smp\right] \Plm.
\end{split}
\end{equation}
Here, the reference radius used is that of Ganymede ($r=r_g$) and 
$P_\ell^m$ denotes the associated Legendre function whose normalization is subject to the Schmidt convention; $\ell_{\text{max}}^I$
(resp.  $\ell_{\text{max}}^E$) is the truncation of the spherical harmonic expansion of internal (resp. external) sources. 
The Gauss coefficients that enter this expression, $(\glm,\hlm)$ for the internal potential and  $(\qlm,\slm)$ for the external potential, 
are expressed in nanoteslas (nT). 
 
We now define the internal and external magnetic fields that interact 
with conductive flow in Ganymede's ocean. Internal (resp. external) implies that the source is located at
$r<r_i$ (resp. $r>r_o$). The internal field is due to Ganymede's dynamo, while the external field is due
to Jupiter's magnetic field. 
\subsection{The external field model}\label{Asec:ExternalModel}
The primary source of the external magnetic field  
is Jupiter's magnetosphere; the temporal variations of this external field are driven by Ganymede's orbital motion around Jupiter, as  
detailed by e.g. \citet{Saur10} and \citet{Seufert11}.
The key factor influencing magnetic field variability is the $9.6^{\circ}$  tilt of Jupiter's internal field 
relative to its spin axis. Due to the rapid rotation of Jupiter (every 10~ hours), which is faster than the satellite's orbital period (of about 7 Earth days), the satellite cyclically passes above and below Jupiter's magnetic equator. 
This positional change induces magnetic variations at satellite location, 
corresponding to the synodic rotation period of Jupiter in the satellite reference frame (10.53~hours for Ganymede). 
Another distinctive period results from the satellite rotation around 
Jupiter, taking 7.16 Earth days for Ganymede. Even in a perfectly axisymmetric 
magnetic field, Ganymede's slightly eccentric orbits introduce time-variable fields in the satellite reference frame. 
In addition, the impact of solar wind on the magnetosphere contributes to 
inducing time-dependent magnetic fields at the satellite's rotation period. 

\begin{figure}[h!]
  \centerline{\includegraphics[width=\textwidth]{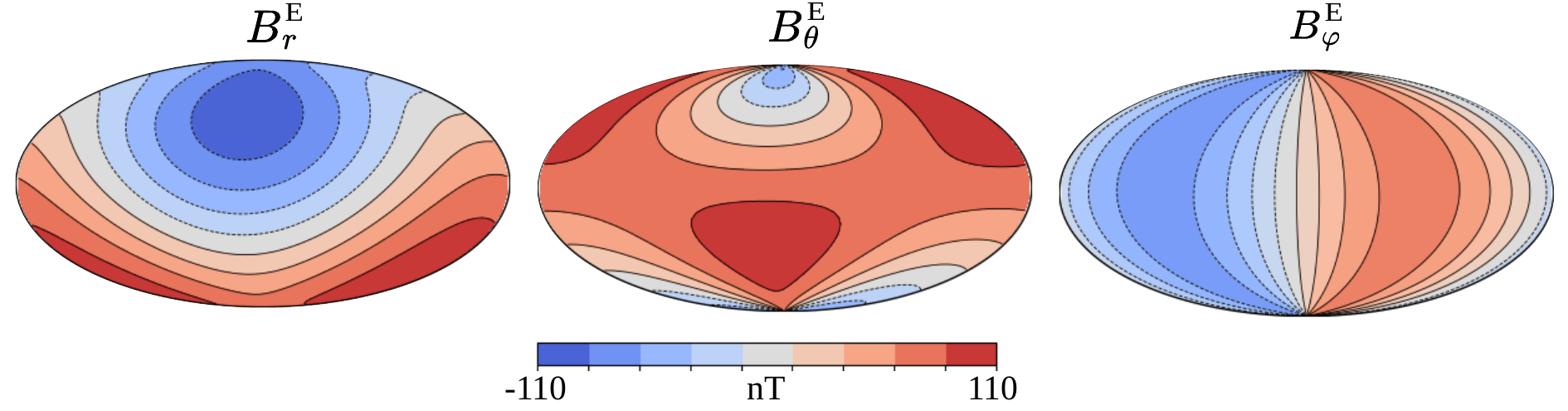}}
    \caption{Maps of the external magnetic field imposed by Jupiter's magnetosphere at 
    Ganymede's reference radius $r_g$, on  August 25, 2023 at 12:00 pm. 
    \label{Afig:ExternalB}
    }
\end{figure}

To model Ganymede's external magnetic field, we use the NAIF-produced SPICE 
kernels \citep{acton18} to obtain the most precise ephemeris data available, as they include the 
orbital perturbations responsible for magnetic oscillation at satellite position. 
The magnetic field computation employs Ganymede-centered right-handed planetocentric (GSPRH) coordinates. 
This coordinate system is constantly rotating with the satellite, and remains fixed to its center. 
The remapping of Jupiter's magnetosphere is performed on a spherical grid consisting of 30 points 
in colatitude and 60 points in longitude at Ganymede's reference radius $r_g$. To include 
low-frequency magnetic field perturbations in our external magnetic model, 
we generated a 1120 hour-long time series of magnetic field maps 
on the sphere, corresponding to 156 Ganymede orbital periods, 
using SPICE generated data from August, 15th, 2023 until September 30th, 2023. 
Figure~\ref{Afig:ExternalB} shows the magnetic field components $B_r^E, B_{\theta}^E, B_{\varphi}^E$ 
(in the 
GSPRH frame) of our external magnetic model at Ganymede's surface $r=r_g$. 
These magnetic field components 
are derived from the revised model of Jupiter's internal magnetic field proposed by \citet{connerney22} and combined 
with the jovian magnetodisc model of \citet{Connerney20}, both of which are available at \url{https://pypi.org/project/con2020/}. 

\begin{figure}[h!]
  \centerline{\includegraphics[width=\textwidth]{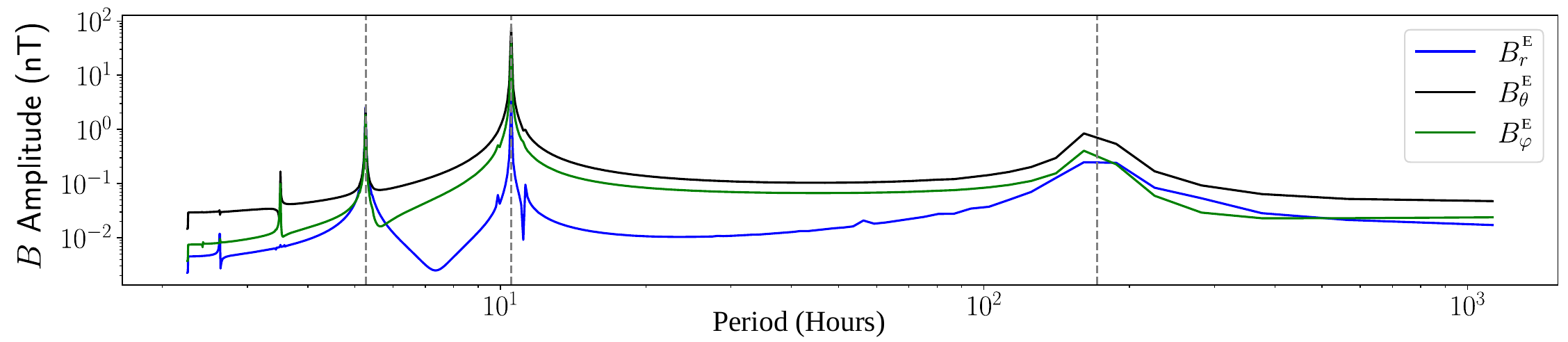}}
    \caption{Frequency spectra of the external magnetic field components 
    in the GSPRH frame computed at Ganymede orbital position. Dashed vertical lines 
    highlight three dominant frequencies $w_1 = 1.083.10^{-5}$~rad.s$^{-1}$ (period \rv{of} 171.57 h), $w_2 = 
    1.656.10^{-4}$~rad.s$^{-1}$ (10.53~h) and $w_3 = 3.312.10^{-4}$~rad.s$^{-1}$ (5.27h). 
    \label{Afig:ExternalBSpectra} }
\end{figure}

In Fig.~\ref{Afig:ExternalBSpectra}, we present the frequency spectrum of each  magnetic field
component time series. These three spectra reveal 
three dominant frequencies related to Ganymede's position relative to Jupiter: $w_1 = 1.083.10^{-5}$ (171.57 h), $w_2 = 1.656.10^{-4}$ (10.53h) and $w_3 = 3.312.10^{-4}$ (5.27h) rad.s$^{-1}$. Peaks at each of those three frequencies are visible on each magnetic component, with varying magnitudes, reaching up to $\approx 76$ nT for $B^E_{\theta}$ at the synodic rotation period of Jupiter (10.53h). 

We estimate the $(\qlm(t), \slm(t))$ 
that best account for $\mathbf{B}^E(t)$ in the usual least-squares sense. Noting that a 
quadrupolar fit ($\ell_{\text{max}}^E = 2$) provides 
a modest increase in accuracy compared with that of a dipolar fit, we favor for the sake of parsimony a dipolar approximation of the
external field ($\ell_{\text{max}}^E = 1$). 
The subsequent incorporation of  $(\qlm(t), \slm(t))$ into our 
simulations is made possible by 
their Fourier expansion 
\begin{equation}\label{AEq:tdpglm}
\begin{split}
\qlm(t) &= \qlmj{0} 
+ 2 \mathrm{Re}\left[ 
\sum_{k=1}^{3} \qlmj{k} \exp i \omega_k t \right], \\
\slm(t) &= 
\slmj{0} 
+ 2 \mathrm{Re}\left[ 
\sum_{k=1}^{3} \slmj{k} \exp i \omega_k t \right],
\end{split}
\end{equation}
where $\mathrm{Re}[x]$ denotes the real part of $x$.
Fourier coefficients for each frequency are given in Table~\ref{Atab:qlmslm} and constitute the external magnetic field model defined at Ganymede's reference radius and used as input in our numerical simulations. 
\begin{table}[h!]
  \caption{\footnotesize 
Fourier coefficients for each frequency are given in nanotesla. 
Frequencies are : $w_1 = 1.083.10^{-5}$, $w_2 = 1.656.10^{-4}$ and $w_3 = 3.312.10^{-4}$ rad.s$^{-1}$.
The real-valued $\qlmj{0}$ and $\slmj{0}$ are the time average of the external Gauss coefficients.}
  \label{Atab:qlmslm}
\centering
 \begin{tabular}{| c | c c c c c c c |} 
 \toprule
   & $\qlmj{0}$ & Re[$\qlmj{1}$] &  Im[$\qlmj{1}]$ &
                  Re[$\qlmj{2}$] &  Im[$\qlmj{2}]$ &  
                  Re[$\qlmj{3}$] &  Im[$\qlmj{3}]$   \\ [0.5ex] 
     \midrule 
  $q^0_1$ & 75.12 & 0.0607 & 0.111 & -0.412 & 1.14 & -0.296 & 1.26\\ [0.5ex] 
  $q^1_1$ & -0.342 & 0.265 & 0.369 & 25.1 & 33.4 & -1.28 & 0.186\\ [0.5ex] 
     \midrule
   & $\slmj{0}$ & Re[$\slmj{1}$] &  Im[$\slmj{1}]$ &
                  Re[$\slmj{2}$] &  Im[$\slmj{2}]$ &  
                  Re[$\slmj{3}$] &  Im[$\slmj{3}]$   \\ [0.5ex] 
     \midrule 
  $s^1_1$ & 0.0124 & -0.0796 & -0.0116 & -7.91 & 1.24 & -0.113 & -0.854\\ [0.5ex] 
 \bottomrule
\end{tabular}
\end{table}

\subsection{The internal dynamo field model}\label{Asec:InternalModel}
The model of Ganymede's magnetic field of internal origin is assumed to be time-independent, implying that the Gauss coefficients $\glm$ and $\hlm$ are considered constant over our simulations time span. 
This assumption can be rooted in the body of work concerned with the 
geomagnetic secular variation $\tau_{\mbox{sv}}$, 
namely the observed time scale of variation of the Earth's dynamo \citep[e.g.][]{christensen04,lhuillier11}. 
Studies of the geodynamo suggest that $\tau_{\mbox{sv}} \sim \tau_d / R_m$, where $\tau_d$ is
the dipole magnetic diffusion time in the core, and $R_m$ is magnetic Reynolds number of the geodynamo. 
Taking 
$\tau_d = r_c^2 / (\pi^2 \lambda)$ with $r_c=3500$~km, the magnetic diffusivity 
$\lambda \approx 1$~m$^2$/s
and 
$R_m \approx 1000$ 
for Earth \citep{roberts13}, we have $\tau_d \approx 40,000 $~yr. Since geomagnetic observations indicate
that 
$\tau_{\mbox{sv}} \mbox{ (Earth) } \approx 400$~yr, we find $ \tau_{\mbox{sv}} \approx 10 \, \tau_d / R_m$. 
Assuming that law holds for Ganymede gives us $\tau_{\mbox{sv}}\mbox{ (Ganymede) }\approx 16$~yr if we take
$R_m\approx 500$, 
one the largest values considered by \citet{christensen15} in his simulations of the Ganymede
 dynamo. In other words, 
it is reasonable to assume that over a simulation time span of a few weeks to months, the large scales
of Ganymede's dynamo are steady. 

Using data acquired by the Galileo and Juno magnetometers during their flybys of Ganymede, \citet{Weber22, plattner23} and \citet{jia25} present models of the internal field after separating out external magnetic contributions from Jupiter's magnetosphere. We assume for our proof-of-concept exercise that
 these coefficients are entirely due to Ganymede's dynamo, neglecting the effect of induction in the ocean. 
The internal field is expressed as the sum of a magnetic dipole and quadrupole, with the Gauss coefficients evaluated at Ganymede’s reference radius, as reported in Table~\ref{tab:IGC}.
\begin{table}[h]
\caption{Dipole and quadrupole Gauss coefficients used to construct 
 internal dynamo models for Ganymede. Coefficients in nanoteslas (nT).}
\begin{tabular}{lrrrrrrrr}
 &    $g_1^0$ & $g_1^1$ & $h_1^1$ & $g_2^0$ & $g_2^1$ & $h_2^1$ & $g_2^2$ & $h_2^2$ \\  \hline 
Quadrupole fit by \citet{Weber22} & -748.3 & 41.1  & 20.8 & 22.5  & 23.3  & 16.5  & -26.8  & -10.6   \\
Model L2A- by \citet{plattner23} & -761.4 & 58.9  &  12.0 & 21.4  & 4.5  & -20.8  & 6.8  & -16.5   \\
MagModel 4 by \citet{jia25}  & -725.6 & 35.7  & 6.9 & 3.8  & 31.9  & 13.5  & -9.8  & -5.4   \\ \hline 
\end{tabular}\label{tab:IGC}
\end{table}

\begin{figure}[h!]
  \centering
  \includegraphics[width=0.8\textwidth]{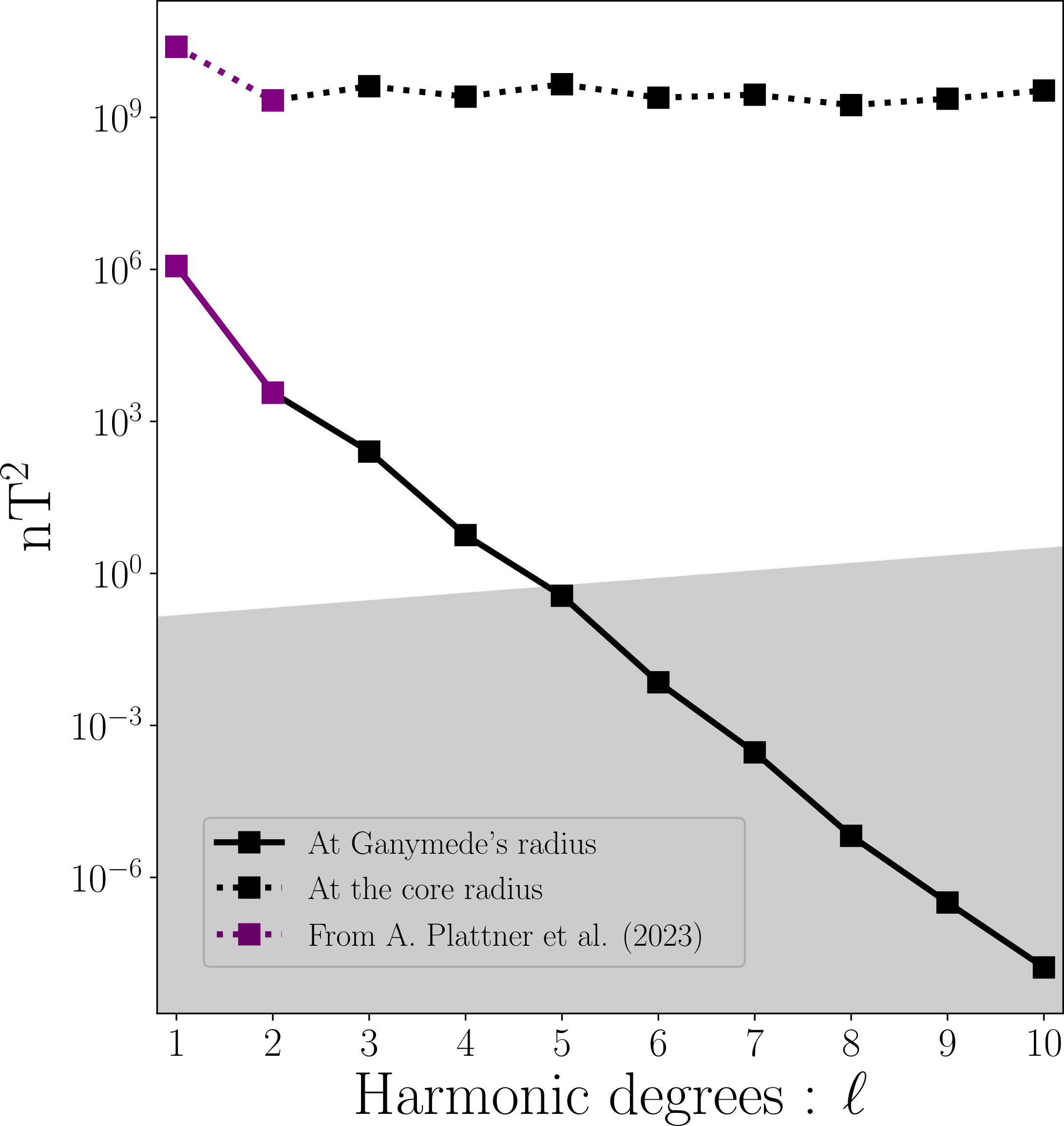}
    \caption{Example magnetic power spectrum of internal origin presented at both Ganymede's reference radius $r_g$ and the core surface $r_c$. These spectra are derived from the the L2A- model by \citet{plattner23}, which prescribe the energy at spherical harmonic degrees $\ell=1$ and $2$.
    Beyond $\ell=2$, we generate a
    flat random power spectrum whose level matches that of the quadrupole $\ell=2$ at the top of the core
     $r=r_c$.
     \rvv{The gray area represents the slope corresponding to the 0.2 nT detection threshold, resulting from the amplification of measurement errors during downward continuation from a spacecraft altitude of 500~km to Ganymede’s surface.}}
     \label{Afig:EMInit}
\end{figure}

Given that the internal Gauss coefficients adhere to the expression of a scalar potential $V^I$ in Eq.~\ref{AEq:VIVE}, we employ the magnetic power spectrum  \citep[e.g.][]{lowes74} that leads to the series,
\begin{equation}\label{AEq:MagPowSpec}
E_{\ell}(r) = (\ell + 1) \left( \frac{r_g}{r} \right)^{2\ell+4} \sum_{m=0}^{\ell} \left[(\glm)^2 + (\hlm)^2 \right].
\end{equation}
Here, $E_{\ell}$ is expressed in nT$^2$ and provides an estimate of the contribution of each spherical harmonic degree $\ell$ to the squared amplitude of the observed magnetic field of internal origin. Figure~\ref{Afig:EMInit} shows the power spectrum, computed from the Gauss coefficients provided by \citet{plattner23}, in purple, for 
$r=r_g$ and $r=r_c$. It appears that 
at $r=r_g$  the dipole ($\ell=1$) 
is significantly more energetic than the quadrupole ($\ell=2$). This remains true at 
the core surface, $r=r_c$. 
Given these dipolar and quadrupolar Gauss coefficients, we now detail our procedure to infer the magnitude of higher harmonic degrees originating from Ganymede's internal dynamo, leading to the \flqq $\mathbf{U} = \mathbf{0}$ \frqq spectrum presented in Figure~4 of the main text. 
 
We extend the internal field model of L2A- by \citet{plattner23}  
by assuming energetic predominance of the magnetic dipole and equipartition of non-dipolar 
magnetic energy
such that for $\ell\geq2$ 
\begin{equation}\label{eq:expectation}
\mathbb{E}\left\{ (l+1) \sum_{m=0}^{\ell} \left[(\glm)^2 + (\hlm)^2 \right] \right\} = E_2 \quad \text{for } \ell > 2.
\end{equation}
Here, $\mathbb{E}{\cdot}$ denotes the expectation operator, and $E_2$ represents the magnetic energy of the quadrupole, as computed \AF{per}  Figure~\ref{Afig:EMInit}
from the internal model L2A- of \citet{plattner23}. Within each degree $\ell$, we further assume that $g_\ell^m$ and $h_\ell^m$ are independent, identically distributed normal random variables with zero mean. If $\text{Var}(g_\ell^m) = \text{Var}(h_\ell^m) = \sigma^2$, then, following the K\"onig-Huygens theorem, we have:
\begin{equation}
\text{Var}(g_\ell^m) = \mathbb{E}\{(g_\ell^m)^2\} - \mathbb{E}\{g_\ell^m\}^2 = \mathbb{E}\{(g_\ell^m)^2\} = \sigma^2,
\end{equation}
with a similar result for $h_\ell^m$. Combined with expression~\eqref{eq:expectation}, this leads to the expression for the variance:
\begin{equation}
\sigma^2 = \frac{E_2}{(\ell+1) (2\ell+1)}.
\end{equation}
Building on these principles, we artificially generate a flat spectrum at the surface of Ganymede's core, 
with energetic values randomly assigned for Gauss coefficients for degrees within the range $3 \leq \ell \leq 10$.
To capture the variability introduced by this randomization, we produce eleven realizations for each reference model—those of \citet{Weber22}, \citet{plattner23}, and \citet{jia25} listed in Table~\ref{tab:IGC}—resulting in a set of $3 \times 11 = 33$ synthetic spectra that form our ensemble of internal magnetic field models. 

\begin{figure}[h!]
  \centering
  \includegraphics[width=\textwidth]{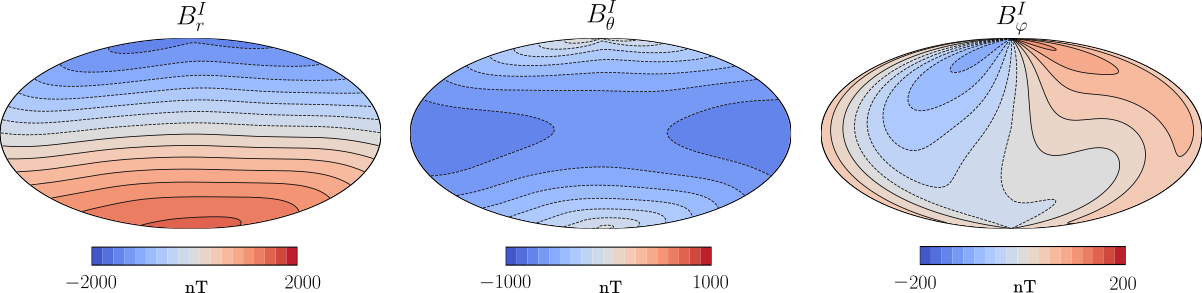}
    \caption{ Maps of the internal magnetic field generated by Ganymede’s core dynamo, evaluated at the reference radius $r_g$ using the internal model L2A- of \citet{plattner23}. 
    \label{Afig:internalB} }
\end{figure}

One such synthetic magnetic power spectrum based on \citet{plattner23}’s L2A- model is shown in Figure~\ref{Afig:EMInit}, both at the surface of the core (black dashed line) and at Ganymede’s reference radius (black solid line). 
Corresponding maps are displayed at Ganymede's \AF{reference} radius in Fig.~\ref{Afig:internalB}. 
Although there is no 
physical ground to justify an exact equipartition between the various degrees of the non-dipole field,
 observations and simulations of the geodynamo (and of Ganymede, see Fig.~8 by \citet{christensen15})
  suggest that this is a reasonable approximation. 
A key property of a flat non-dipole magnetic power spectrum at the top of the core $r=r_c$, 
is that small-scales (large degrees) are strongly attenuated with upward continuation.  
This results in a steep slope in the magnetic power spectrum observed at Ganymede's reference radius, 
corresponding to a $(r_c/r_g)^{2\ell}$ law for its non-dipole part. 
This geometric attenuation is crucial to unequivocally distinguish magnetic contributions from the ocean and the core at Ganymede's surface, specifically at large degrees $\ell$.

\rvv{As shown in Figure~\ref{Afig:EMInit}, attenuation beyond $\ell = 5$ significantly reduces the detectability of the internal dynamo signal, whose magnetic amplitude approaches the 0.2~nT detection threshold of JUICE’s magnetometers (indicated by the gray area for a spacecraft altitude of 500~km). This threshold follows \cite{brown19}, who report a peak-to-peak sensitivity of 0.2~nT for measurements acquired within the $0.3~\mu\mathrm{Hz}$ -- $100~\mathrm{mHz}$ frequency band, corresponding to periods from approximately 39~days to 10~s. 
A typical horizontal 
magnetic length scale $L$ can be estimated as 
a function of SH degree $\ell$ by 
$L \approx \pi r/\ell$ \citep[see][Sec.~3.6.3]{backus1996foundations}.  
At a circular orbit at 500~km altitude ($r \approx 3131~\mathrm{km}$), 
JUICE’s orbital speed is $v \approx 1.5~\mathrm{km/s}$. 
Over the shortest period of the frequency band 
discussed above (namely 10~s), the spacecraft is therefore
expected to travel 
$\sim 15~\mathrm{km}$. 
Requiring at least twice this distance to resolve a spatial scale gives 
$L_\mathrm{min} \sim 30~\mathrm{km}$, corresponding to a maximum spherical harmonic degree of $\ell \sim 328$ at a precision of 0.2~nT. Since our spectra extend only to $\ell = 10$, the associated spatial scales are well above the minimum length scale resolvable at the 0.2~nT detection threshold.
}

\section{Magnetic boundary conditions}\label{Asec:BCs}
In order to model the interaction of the internal and external magnetic fields defined above  
with Ganymede's ocean, appropriate 
boundary conditions at $r=r_i$ and $r=r_o$ are in order. 
As illustrated in Fig.~\ref{fig:coupeGany}, Ganymede's ocean is sandwiched between 
two electrically insulating layers. 
 Inside a conducting layer, the magnetic field is governed by 
the induction equation and is decomposed into a poloidal field  and a toroidal field, following
\[
\begin{split}
    \mathbf{B} &= \mathbf{B}^P + \mathbf{B}^T,\\
               &= \bnabla \times (\bnabla \times B^P \mathbf{e}_r) + \bnabla \times B^T \mathbf{e}_r,
\end{split}
\]
On the other hand, in an insulating layer, the magnetic field is a potential field, $\mathbf{B} = - \bnabla V$. In the following, $V^{-}$ (resp. $V^+$) will denote the magnetic potential below (resp. above) the ocean. 

The three components of the magnetic field are continuous across an interface 
of finite conductivity jump. 
Consequently, the toroidal scalar $B^T$ vanishes at the boundaries and 
the poloidal scalar $B^P$ and its radial derivative are subjected to \citep[e.g.][Appendix III]{chandra61}
\begin{equation}\label{AEq:GBCo}
\begin{split}
-\frac{1}{r^2}\bnabla_H^2 B^P &= -\partial_r V^\pm, \\
\bnabla_H \partial_r B^P &= - \bnabla_H V^\pm,\\  
\end{split}
\end{equation}
where $\bnabla_H$ and $\bnabla_H^2$  are the horizontal gradient and Laplace operators on the unit sphere, respectively, and 
we understand that $V^\pm=V^-$ at $r=r_i$ and $V^\pm=V^+$ at $r=r_o$. 

In the \texttt{MagIC} code, the scalar fields $V^\pm$ and $B^P$ are expanded using spherical harmonics, 
\begin{equation}\label{AEq:VBp}
\begin{split}
V^\pm\rtp &= \sum_{\ell=1}^{\lmax} \sum_{m=-\ell}^{\ell} \vlm^\pm(r) \ylm \tp,\\
B^P\rtp &= \sum_{\ell=1}^{\lmax} \sum_{m=-\ell}^{\ell} \blm(r) \ylm \tp,\\
\end{split}
\end{equation}
where the spherical harmonics $\ylm(\theta, \varphi)$ of degree $\ell$ and order $m$ are complex-valued, fully normalised eigenfunctions of  
$\bnabla_H^2$.  

Within an insulating layer, the radial dependence of the scalar potential $\vlm^\pm(r)$ 
enables separation of contributions from internal and external sources. 
In the mantle, using $r_M$ as reference radius ($ r_c\leq r_M\leq r_i$), we write accordingly  
\begin{equation}
\vlm^-(r) = r_M \left[ \vilmm \left( \frac{r_M}{r} \right)^{\ell +1} + \velmm \left( \frac{r}{r_M} \right)^{\ell}\right],
\end{equation}
where $\vilmm$ and $\velmm$ are complex-valued coefficients for the internal and external sources, respectively. 
In Ganymede's mantle and deep ice layer, the $\vilmm$ are due to Ganymede's dynamo, that operates at $r\leq r_c$, while the 
$\velmm$ are due to currents in the ocean and beyond, 
located at $r\geq r_i$. We now make use of the two equations \rv{in} Eq.~\eqref{AEq:GBCo}, written at $r=r_i$ to 
eliminate the unwanted $\velmm$, to obtain the following relationship  
\begin{equation}
\left( \frac{\mathrm{d}}{\mathrm{d}r} -\frac{\ell+1}{r_i} \right)  \blm  = -r_i\frac{2\ell+1}{\ell} \left(\frac{r_M}{r_i}\right)^{\ell+2} \vilmm \mbox{ at } r=r_i. 
\label{eq:magBCinnerMagic}
\end{equation}
The same reasoning can be applied to $V^+$, the magnetic potential in the ice cap and beyond. Using 
$r_s\geq r_o$ as reference radius, we have 
\begin{equation}
\vlm^+(r) = r_s \left[ \vilmp \left( \frac{r_M}{r} \right)^{\ell +1} + \velmp \left( \frac{r}{r_M} \right)^{\ell}\right],
\end{equation}
where $\vilmp$ and $\velmp$ are complex-valued coefficients for the internal and external sources, respectively. 
External sources correspond here to the time-dependent Jovian field described above, while internal sources 
combine the dynamo field and the field induced in the ocean. 
It is that latter unwanted component that we eliminate this time, 
which leads to the following condition at $r=r_o$ 
\begin{equation}
\left( \frac{\mathrm{d}}{\mathrm{d}r} +\frac{\ell}{r_o} \right)  \blm  = -r_o\frac{2\ell+1}{\ell+1}
\left(\frac{r_o}{r_s}\right)^{\ell-1} \velmp \mbox{ at } r=r_o.
\label{eq:magBCouterMagic}
\end{equation}
It is now a matter of expressing  $\velmp$ and $\vilmm$ in
Eqs. \eqref{eq:magBCinnerMagic} and \eqref{eq:magBCouterMagic} in 
terms of the Gauss coefficients $(\qlm,\slm)$ and $(\glm,\hlm)$, respectively, keeping in mind 
that these were introduced using $r_g$ as reference radius, and that they rely on Schmidt-normalized real-valued spherical harmonics. We find 
\begin{align}
\left( \frac{\mathrm{d}}{\mathrm{d}r} -\frac{\ell+1}{r_i} \right)  \blm  = 
-\frac{r_i}{\ell} \left( \frac{r_g}{r_i}\right)^{\ell+2}  \sqrt{\frac{4 \pi (2\ell+1)}{2-\delta_{m0}}}  (\glm - i\hlm)
& \mbox{ at } r=r_i. \label{eq:bc_inner}\\
\left( \frac{\mathrm{d}}{\mathrm{d}r} +\frac{\ell}{r_o} \right)  \blm  =  
- \frac{r_o}{\ell+1} \left(\frac{r_o}{r_g}\right)^{\ell-1} \sqrt{\frac{4 \pi (2\ell+1)}{2-\delta_{m0}}} (\qlm - i\slm) & \mbox{ at } r=r_o, \label{eq:bc_outer}
\end{align}
where $\delta$ is the Kronecker symbol. 
Both conditions are implemented and enforced within {\tt MagIC} upon advancing the $\blm$'s in time. Let us stress here that in our approach, we assume that 
 the $\glm$ and $\hlm$ are steady, while the $\qlm$ and $\slm$ are time-dependent. 
From the numerical standpoint, the treatment of the time-dependency of the right-hand side of the non-homogeneous Robin condition at $r=r_o$ is fully implicit. 
 We verified our implementation considering magnetic diffusion through a motionless ($\mathbf{U}=\mathbf{0}$) homogeneous
 ocean, using the analytical solution detailed in \citet{zimmer00,Saur10}. 

 \rv{In practice, 
 \texttt{MagIC} has an operation mode (control integer parameter \texttt{mode=2}) under which the induction equation 
is the only equation advanced in time. 
The time-dependent external and constant internal imposed magnetic fields 
enter the problem solved 
by means of Equations 
 \eqref{eq:bc_inner}-\eqref{eq:bc_outer}, which take into account the different normalization used in 
 $\texttt{MagIC}$ and geomagnetism. 
 They are implemented in the \texttt{updateB} routine of the
 \texttt{updateB\_mod} module, are dealt with fully implicitly, and are directly
 expressed in nT, since the induction equation is linear in the magnetic field $\mathbf{B}$, which implies that 
 no rescaling is necessary. }

\section{Gauss coefficients from numerical simulations}
\label{sec:cooking}
Given that MagIC simulations deliver the magnetic field at the top of the conductive ocean $r_o$, we need to perform an upward continuation of the induced magnetic field through the ice layer to derive magnetic power spectra at Ganymede's radius. At this radius, the radial component of the magnetic field reads,
\begin{equation}\label{AEq:Brrg}
\begin{split}
B_r(r,\theta,\phi,t) &= -\frac{\partial V}{\partial r} \\
 &= \sum_{\ell=1}^{\ell_{\text{max}}} \sum_{m=0}^{\ell} (\ell + 1) \left( \frac{r_g}{r} \right)^{\ell+2} \left[\glm(t) \cmp + \hlm(t) \smp \right] \Plm\\
 &- \sum_{\ell=1}^{\ell_{\text{max}}} \sum_{m=0}^{\ell} \ell  \left( \frac{r}{r_g} \right)^{\ell-1} \left[\qlm(t) \cmp + \slm(t)\smp \right] \Plm.
\end{split}
\end{equation}
Here, $V$ represents the scalar potential in an insulating layer as given by Eqs.~\ref{AEq:VIVE}. 
At the top of the ocean, the radial magnetic field is continuous across the interface. Since 
MagIC provides internal Gauss coefficients for a reference radius $r_o$, denoted by $\glmmag$ and $\hlmmag$, the radial magnetic field can also be expressed as,
\begin{equation}\label{AEq:Brro}
    B_r(r,\theta,\phi,t) = \sum_{\ell=1}^{\ell_{\text{max}}} \sum_{m=0}^{\ell} (\ell + 1) \left( \frac{r_o}{r} \right)^{\ell+2} \left[ \glmmag(t) \cmp + \hlmmag(t) \smp \right] \Plm.
\end{equation}
Equality between equations Eqs~\eqref{AEq:Brrg} and \eqref{AEq:Brro} at $r=r_o$ implies,
\begin{equation}
\begin{split}
\glm(t) &=  \left( \frac{r_o}{r_g} \right)^{\ell+2} \left[ \glmmag(t) + \frac{\ell}{\ell+1}  \left( \frac{r_o}{r_g} \right)^{\ell-1} \qlm(t) \right],\\
\hlm(t) &=  \left( \frac{r_o}{r_g} \right)^{\ell+2} \left[ \hlmmag(t) + \frac{\ell}{\ell+1}  \left( \frac{r_o}{r_g} \right)^{\ell-1} \slm(t) \right].
\end{split}
\end{equation}
Using $r=r_g$ as the reference radius, these Gauss coefficients are used to derive magnetic power spectra of internal origin following \citep[][\S4.4]{backus1996foundations}
\begin{equation}
E_{\ell}(t) = (\ell + 1)  \sum_{m=0}^{\ell} 
\left\{
\left[\glm(t)\right]^2 + \left[\hlm(t)\right]^2
\right\}.
\end{equation}
The time averages of the spectra so obtained are presented in Fig.~4 in the main text.

\section{The Poloidal-Toroidal Representation} \label{Asec:torpol}
Our numerical approach involves decomposing the divergence-free magnetic field $\mathbf{B}$ into poloidal ($\mathbf{B}^P$) and toroidal ($\mathbf{B}^T$) components by introducing the scalar potentials $B^P(r,\theta,\phi,t)$ and $B^T(r,\theta,\phi,t)$, such that,
\begin{equation}\label{AEq:PolTor}
\begin{split}
    \mathbf{B} &= \mathbf{B}^P + \mathbf{B}^T,\\
               &= \bnabla \times (\bnabla \times B^P \mathbf{e}_r) + \bnabla \times B^T \mathbf{e}_r,
    \end{split}
\end{equation}
The poloidal and toroidal potentials, $B^P_{\ell m}$ and $B^T_{\ell m}$, are expanded in spherical harmonics. The total magnetic energy per unit volume within the conducting ocean is then given by,
\begin{equation}
\begin{split}
E_\ell &= E_\ell^P + E_\ell^T,\\
&= \frac{1}{\pi(r_o^2-r_i^2)}\int_{r_i}^{r_o}  \sum_{m=0}^\ell \frac{\ell(\ell+1)}{\mu} \left[ \frac{\ell(\ell +1)}{r^2} |B^P_{\ell m}|^2 + \left| \frac{\partial B^P_{\ell m} }{\partial r} \right|^2 + |B^T_{\ell m}|^2 \right] \mathrm{d}r
\end{split}
\end{equation}
where $E_\ell^P$ and $E_\ell^T$ represent the poloidal and toroidal magnetic energy contributions, respectively. Here, the integration is over the conducting shell disc bounded by internal radius $r_i$ and external radius $r_o$, and $\mu$ is the magnetic permeability \citep[see \S~10.6][]{glatzmaier2013}.

\section{Ocean-induced magnetic field maps} \label{Asec:OIMFM}
\begin{figure}[h!]
  \centering
  \includegraphics[width=\textwidth]{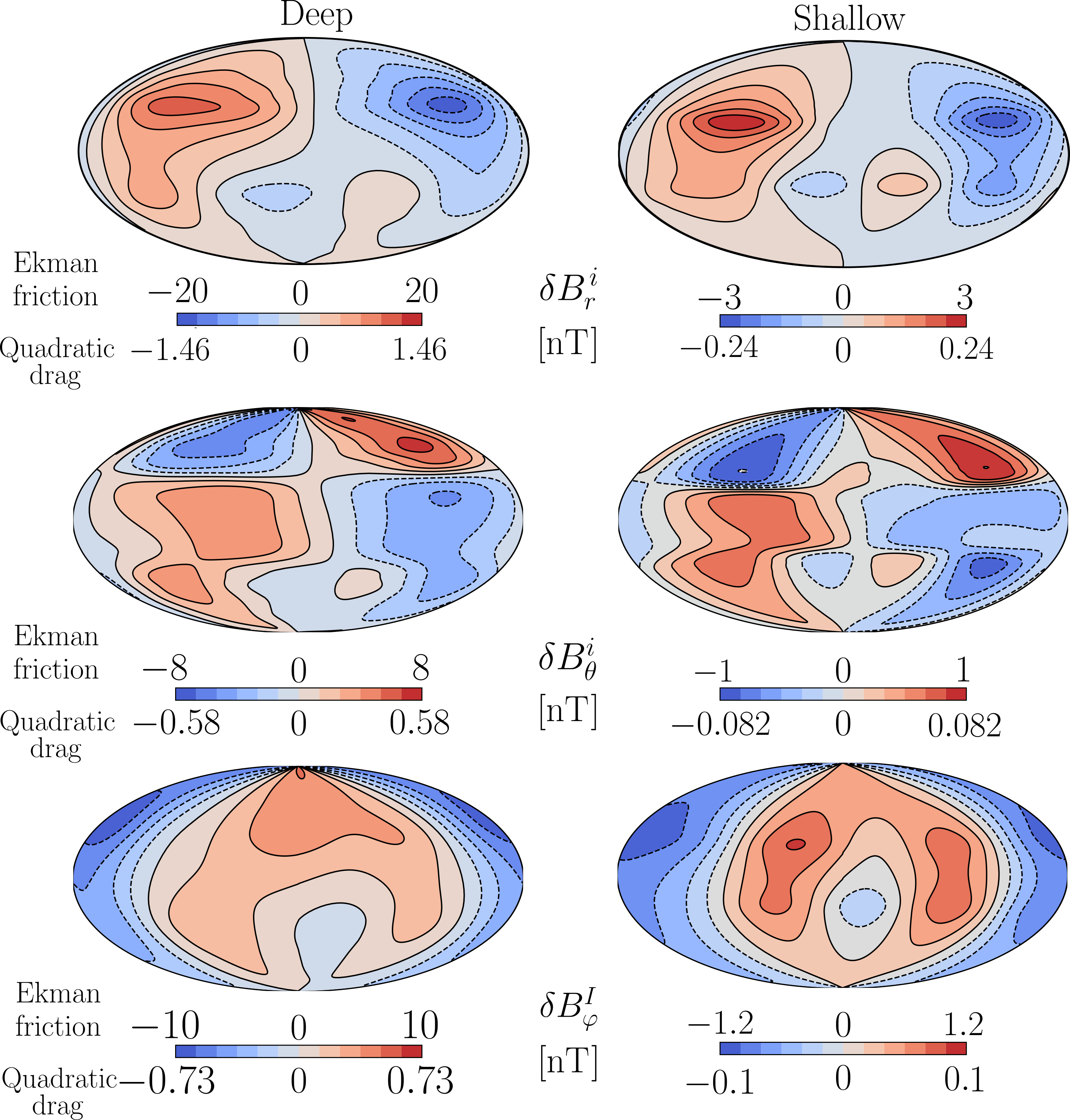}
    \caption{ \rv{Hammer maps of the three components of the ocean-induced magnetic field at the ocean surface ($r = r_o$):
$\delta B_r^i = \hat{\mathbf{r}} \cdot \mathbf{B}_{\mathrm{mi}}$,
$\delta B_\theta^i = \hat{\boldsymbol{\theta}} \cdot \mathbf{B}_{\mathrm{mi}}$,
and $\delta B_\varphi^i = \hat{\boldsymbol{\varphi}} \cdot \mathbf{B}_{\mathrm{mi}}$.
The amplitudes reported rest on a velocity scaling based either on the Ekman friction assumption (top)
or on the quadratic drag assumption (bottom). See main text for details. 
}
    \label{Afig:OIMFM} }
\end{figure}

\rv{Figure~\ref{Afig:OIMFM} shows maps of the magnetic field components—radial, meridional, and azimuthal—induced by oceanic flow at the ocean surface ($r = r_o$), defined as $\delta B_r^i = \hat{\mathbf{r}} \cdot \mathbf{B}_{\mathrm{mi}}$, $\delta B_\theta^i = \hat{\boldsymbol{\theta}} \cdot \mathbf{B}_{\mathrm{mi}}$, and $\delta B_\varphi^i = \hat{\boldsymbol{\varphi}} \cdot \mathbf{B}_{\mathrm{mi}}$. While the deep- and shallow-ocean scenarios seem to yield nearly identical patterns, a deeper ocean slightly smooths non-axisymmetric features, making them less apparent than in the shallow case. 
The symmetry of the magnetic pattern is preserved across both scenarios, contrasting with the anti-correlated patterns reported by \cite{kvorka26}, which depend on the latitudinal distribution of the jets.
This difference likely arises from the no-slip boundary conditions in our hydrodynamic simulations (unlike the free-slip boundaries in \cite{kvorka26}), which prevent jet formation inside the tangent cylinder and produce a two-jet configuration outside it in both ocean cases. Consequently, in our induction setup, the flow structure mainly affects the magnetic field strength while maintaining very similar spatial patterns across radial, meridional, and azimuthal components. A more systematic exploration of a wider range of flow configurations is left for future work and may yield a richer variety of magnetic field patterns.}

\clearpage
\nocite{acton18,backus1996foundations,chandra61,christensen04,glatzmaier2013,lhuillier11}
\bibliography{agusample}

\end{document}